\documentclass[conference]{IEEEtran}
\IEEEoverridecommandlockouts
\usepackage{cite}
\usepackage{amsmath,amssymb,amsfonts}
\usepackage{algorithmic}
\usepackage{graphicx}
\usepackage{textcomp}
\usepackage{xcolor}

\usepackage{tikz}
\usepackage{pgfplots}
\usepackage{subcaption}
\usepackage{multirow}
\usepackage{bm}
\usepackage{hyperref}
\usepackage{url}
\usepackage{booktabs}

\newcommand{\img}{\mathbf{x}_t}
\newcommand{\sysout}{\hat{\mathbf{x}}_t}
\newcommand{\latent}{\mathbf{y}}
\newcommand{\qlatent}{\hat{\latent}}
\newcommand{\synthparam}{\bm{\theta}}
\newcommand{\armparam}{\bm{\psi}}
\newcommand{\upparam}{\bm{\upsilon}}
\newcommand{\synth}{f_{\synthparam}}
\newcommand{\arm}{f_{\armparam}}
\newcommand{\upsample}{f_{\upparam}}
\newcommand{\pred}{\bm{\tilde{x}}_t}
\newcommand{\predictionmode}{\bm{\alpha}_t}
\newcommand{\bidirectionalweight}{\bm{\beta}_t}
\newcommand{\residue}{\bm{r}_t}
\newcommand{\refone}{\hat{\mathbf{x}}_{\mathrm{ref}_1}}
\newcommand{\reftwo}{\hat{\mathbf{x}}_{\mathrm{ref}_2}}
\newcommand{\vtone}{\hat{\mathbf{v}}_{1}}
\newcommand{\vtwo}{\hat{\mathbf{v}}_{2}}

\newcommand{\motionrefone}{\mathbf{v}_{1}}
\newcommand{\motionreftwo}{\mathbf{v}_{2}}

\definecolor{darkpastelpurple}{rgb}{0.59, 0.44, 0.84}
\definecolor{myred}{HTML}{E76F51}
\definecolor{myblue}{HTML}{376996}
\definecolor{mygreen}{HTML}{2A9D8F}
\definecolor{mypurple}{HTML}{822E81}
\definecolor{mywhite}{HTML}{f0dfbb}

\def\BibTeX{{\rm B\kern-.05em{\sc i\kern-.025em b}\kern-.08em
    T\kern-.1667em\lower.7ex\hbox{E}\kern-.125emX}}
\begin{document}

\title{Improved Encoding for Overfitted Video Codecs}

\author{\IEEEauthorblockN{Thomas Leguay$^{\dagger\star}$, Théo Ladune$^{\dagger}$, Pierrick Philippe$^{\dagger}$, Olivier Déforges$^{\star}$}
\IEEEauthorblockA{$^{\dagger}$Orange Innovation, France, \texttt{\small firstname.lastname@orange.com} \\
$^{\star}$Univ Rennes, INSA Rennes, CNRS, IETR (UMR 6164), France, \texttt{\small firstname.lastname@insa-rennes.fr}
}
}

\maketitle

\begin{abstract}
    Overfitted neural video codecs offer a decoding complexity orders of
    magnitude smaller than their autoencoder counterparts. Yet, this low
    complexity comes at the cost of limited compression efficiency, in part due
    to their difficulty capturing accurate motion information. This paper
    proposes to guide motion information learning with an optical flow
    estimator. A joint rate-distortion optimization is also introduced to
    improve rate distribution across the different frames. These contributions
    maintain a low decoding complexity of 1300 multiplications per pixel while
    offering compression performance close to the conventional codec HEVC
    and outperforming other overfitted codecs. This work is made
    open-source at \url{https://orange-opensource.github.io/Cool-Chic/}.
    \end{abstract}


\section{Introduction \& Related works}

Video compression is dominated by the conventional codecs H.264/AVC, H.265/HEVC and
H.266/VVC \cite{avc, hevc, vvc}. One of their main principles is to optimize the
video RD (rate-distortion) cost during encoding to find the best decoder parameters.
Successive generations of codecs offer additional decoding options,
increasing compression performance while maintaining low decoder
complexity.

Recently, autoencoder-based codecs
\cite{dcvc-fm,DBLP:conf/cvpr/LiLL23,9941493,aivc} have leveraged neural networks
to automatically design (learn) a codec and now challenge the state-of-the-art
conventional codec VVC. They do not rely on encoding-time RD optimization of the
decoder parameters. This is delegated to an offline training stage where all the
parameters are learned and set once and for all. Consequently, their decoder
must provision parameters to deal with all possible signals resulting in an
important complexity, up to a million multiplications per decoded pixel
\cite{dcvc-fm, DBLP:conf/cvpr/LiLL23}. This makes their practical use less
attractive, as they are expected to run on a variety of low-power devices.

Overfitting-based codecs \cite{coolchic,mmsp,blard,c3,nerv,hinerv} bridge the
gap between these two paradigms by introducing encoding-time RD optimization
into a learned codec. To this end, a decoder is learned for each image
or video. The successive refinements of the overfitted image codec Cool-chic
\cite{coolchic,mmsp,blard,c3} demonstrate that a neural decoder with 2000
multiplications per decoded pixel is competitive with VVC for image coding.

Several works extend Cool-chic to video coding \cite{c3, coolchicvideo}.
Although these methods guarantee low-complexity decoding, they are far from
challenging conventional methods in terms of rate-distortion performance. These
approaches differ in their exploitation of temporal redundancies. C3 \cite{c3}
extends the spatial entropy module to also account for temporal
redundancies. Cool-chic video \cite{coolchicvideo} relies on residual coding and
motion compensation to leverage temporal redundancies. Yet, Cool-chic video
struggles to capture challenging motion leading to degraded temporal prediction
and limiting its compression performance.

This work improves Cool-chic video \cite{coolchicvideo} by guiding the motion
information learning with a pre-trained optical flow estimator, ensuring
high-quality motion data while maintaining low decoder complexity. An additional
joint optimization stage is introduced to better allocate the rate across
frames. These contributions result in a lightweight video codec (1300
multiplications per decoded pixel) approaching HEVC compression performance and
outperforming other overfitted codecs.

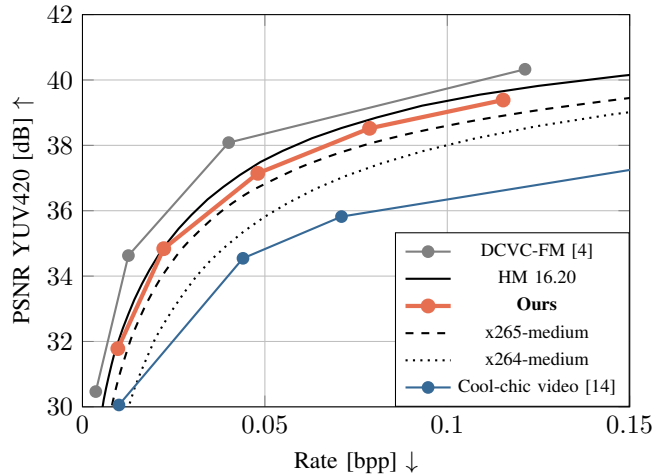
\begin{figure}[t]
    \begin{tikzpicture}
        \begin{axis}[
            grid= both,
            xlabel = {\small Rate [bpp] $\downarrow$},
            ylabel = {\small PSNR YUV420 [dB] $\uparrow$ } ,
            xmin = 0., xmax = 0.15,
            ymin = 30, ymax = 42,
            ylabel near ticks,
            xlabel near ticks,
            width=\linewidth,
            height=6.8cm,
            xtick distance={0.05},
            ytick distance={2},
            ylabel style={yshift=-2pt},
            xticklabel style={
                /pgf/number format/fixed,
                /pgf/number format/precision=2
            },
            minor y tick num=0,
            minor x tick num=0,
            legend style={at={(1.0,0.)}, anchor=south east},
        ]

            \addplot[thick, gray, mark=*, mark options={solid}] table [x=rate_bpp,y=psnr_db] {data/MPEG-B/dcvc.tsv};
            \addlegendentry{\scriptsize DCVC-FM \cite{dcvc-fm}}

            \addplot[thick,  black, mark=none, mark options={solid}] table [x=rate_bpp,y=psnr_db] {data/MPEG-B/hm.tsv};
            \addlegendentry{\scriptsize HM 16.20}

            \addplot[ultra thick,  myred, mark=*, mark options={solid}] table [x=rate_bpp,y=psnr_db] {data/MPEG-B/ours.tsv};
            \addlegendentry{\scriptsize \textbf{Ours}}

            \addplot[thick, black, mark=none, smooth, dashed] table [x=rate_bpp,y=psnr_db] {data/MPEG-B/x265.tsv};
            \addlegendentry{\scriptsize x265-medium}

            \addplot[thick, dotted, black, mark=none, mark options={solid}] table [x=rate_bpp,y=psnr_db] {data/MPEG-B/x264.tsv};
            \addlegendentry{\scriptsize x264-medium}

            \addplot[thick,  myblue, mark=*, mark options={solid}] table [x=rate_bpp,y=psnr_db] {data/MPEG-B/coolchic.tsv};
            \addlegendentry{\scriptsize Cool-chic video \cite{coolchicvideo}}

        \end{axis}
    \end{tikzpicture}
    \caption{Random access rate-distortion graphs. HEVC class B. PSNR computed in YUV420 domain.}
    \label{fig:rd-graph-ra:hevc-b}
\end{figure}

\begin{table}[t]
    \centering
    \caption{BD-rates vs. HEVC (HM 16.20) on the 9 first frames of the HEVC sequences. Distortion in YUV420.}
    \begin{tabular}{c|cccc}
                                            & \multicolumn{4}{c}{Random access BD-rate vs. HM 16.20 {[}\%{]} $\downarrow$ } \\ 
                                            & Class B      & Class C       & Class E       & Average  \\    
    \midrule
    \midrule
    Ours                                    & \textbf{4.6} & \textbf{12.4} & \textbf{10.8} & \textbf{9.3} \\ 
    x265-medium                             & 30.4         & 31.5          & 46.7          & 36.2     \\    
    x264-medium                             & 85.7         & 48.6          & 83.2          & 72.5     \\    
    Cool-chic video \cite{coolchicvideo}    & 118.8        & 100.3         & 105.2         & 108.1    \\    
    \end{tabular}
    \label{tab:all-bd-rate}
\end{table}

\section{Proposed decoding scheme}

\begin{figure*}[t]
    \centering
    \includegraphics[width=\linewidth]{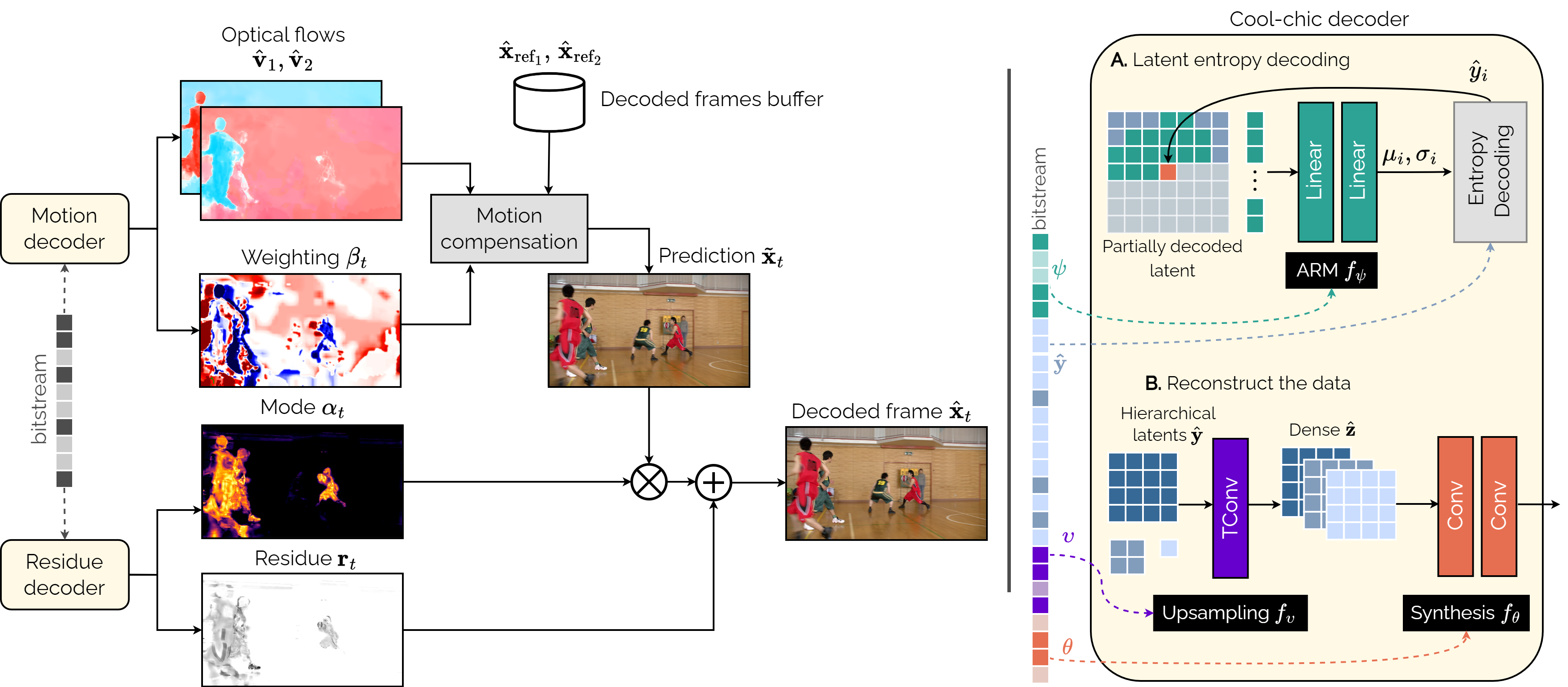}
    \caption{Decoding pipeline of an inter frame with two references. Cool-chic
    decoder diagram is from Blard \textit{et al.} \cite{blard}.}
    \label{fig:frame-decoding}
\end{figure*}

Figure \ref{fig:frame-decoding} illustrates the proposed decoding scheme for a
$H \times W$ frame $\sysout$ with two references $(\refone, \reftwo)$. Each
frame is decoded with their own pair of Cool-chic decoders \cite{coolchic}: one
for the motion and the other for the residue. A Cool-chic decoder is made of 3
NNs (neural networks) and reconstructs data from a latent representation in two
steps. \textit{i)} the auto-regressive NN $\arm$ drives an entropy decoder to
obtain a hierarchical latent representation $\qlatent_{t}$ from the bitstream.
\textit{ii)} the upsampling NN $\upsample$ converts the hierarchical latents
into a dense representation, fed to the synthesis NN $\synth$ to obtain the
decoded data.

The motion decoder outputs two optical flows $\vtone$, $\vtwo$ and a
bi-directional prediction weighting $\bidirectionalweight \in [0, 1]^{H \times
W}$. The flows are pixel-wise motion fields indicating each pixel displacement
between the frame and its references. They are used to obtain a temporal
prediction $\pred$ via a weighted bilinear warping:
\begin{equation}
    \pred
    = \bidirectionalweight \mathrm{warp}(\refone, \vtone)
    + (1 - \bidirectionalweight) \mathrm{warp}(\reftwo, \vtwo).
\end{equation}

The residue decoder outputs a coding mode $\predictionmode \in [0, 1]^{H \times
W}$ and a residue $\residue$. The coding mode is a pixel-wise continuous
weighting masking out the prediction if it is counterproductive. The residue
conveys the non-predicted information. It is added to the masked prediction to
obtain the decoded frame $\sysout$:
\begin{equation}
    \sysout = \predictionmode \pred + \residue.
\end{equation}
A fixed downsampling (\textit{e.g.} bilinear) of the decoded frame is
used to obtain a frame in the YUV420 domain. The proposed decoding process also
considers frames with 1 reference (P-frame) and no reference
(I-frame). For P-frames, only a single optical flow is used and
$\bidirectionalweight = 1$. I-frames are entirely conveyed via the residue
$\residue$ \textit{i.e.} $\predictionmode = 0$.

The key difference with prior works is the two separated Cool-chic decoders for
each frame. Prior methods have a unique Cool-chic decoder for each frame
\cite{coolchicvideo} or shared between several frames \cite{c3}.

\section{Encoding method}

\subsection{Joint optimization of successive frames}
\label{sec:joint-optimization}

\begin{figure}[b]
    \centering
    \includegraphics[width=0.7\linewidth]{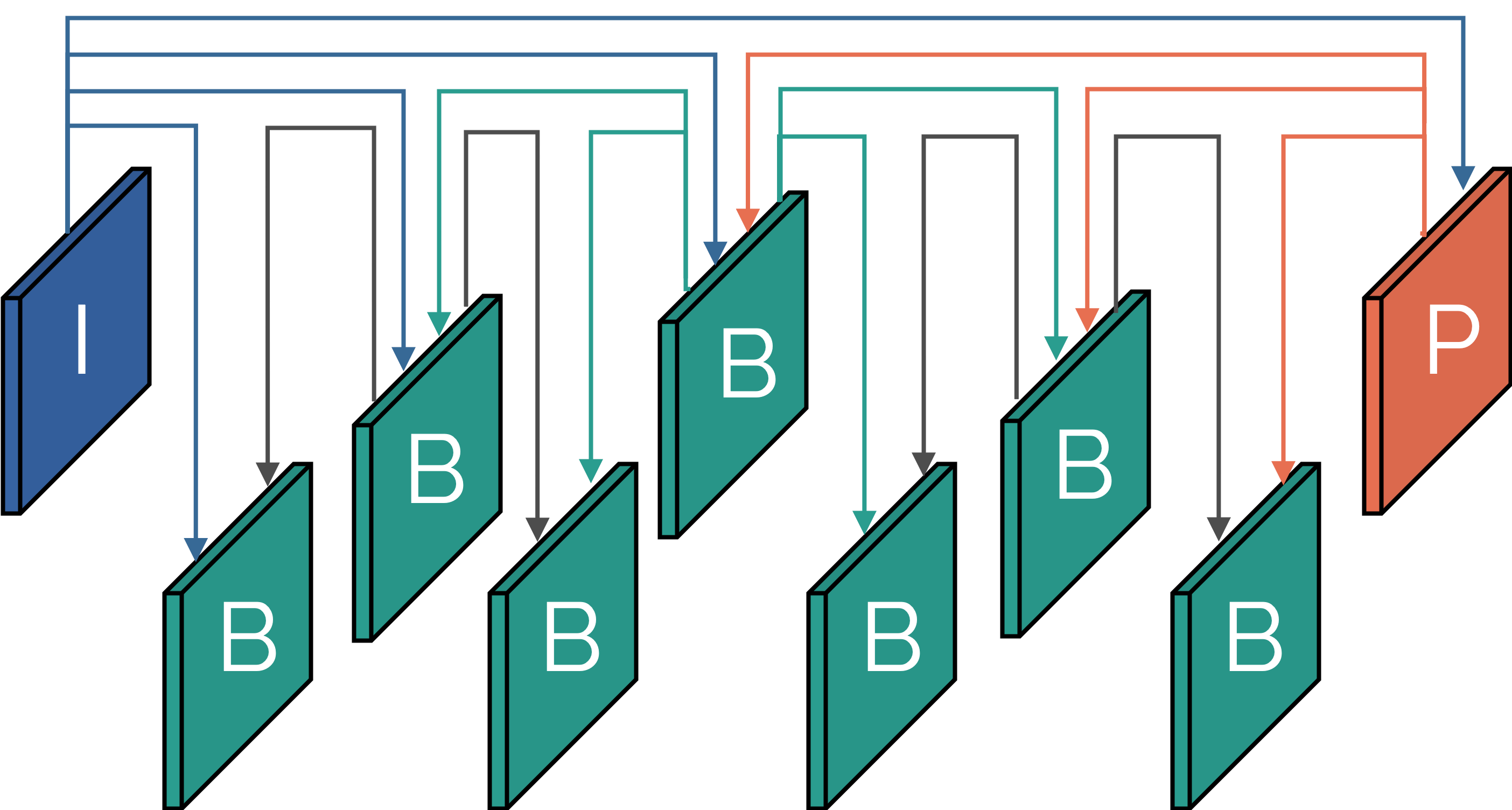}
    \caption{Random access coding configuration.}
    \label{fig:gop:ra}
\end{figure}

Videos are encoded by overfitting the decoder parameters and the latent
representation of each frame to minimize the video RD cost. One key issue to
obtain competitive performance is to properly allocate the rate between the
different video frames. For instance, in the 9-frame hierarchical structure
presented in Fig. \ref{fig:gop:ra}, the I-frame and the middle B-frame are used
as reference for 4 frames whereas some other B-frames never serve as reference.
Consequently, it is usually more effective to allocate more rate to the frames
often used as reference. If such frames exhibit less distortion, 
then the subsequent predictions also present less distortion allowing for 
a lower overall rate-distortion cost.

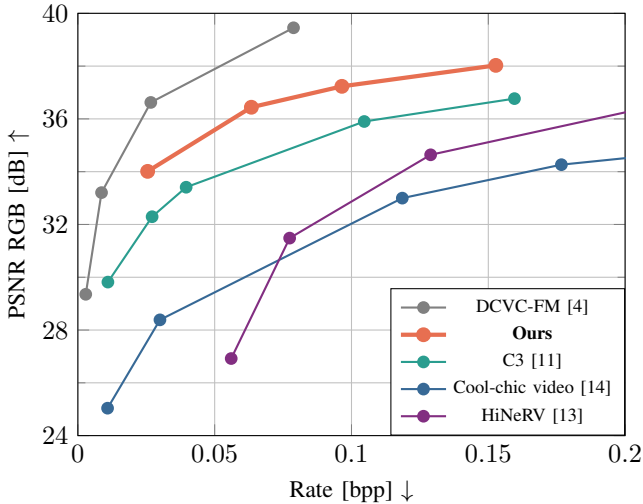
\begin{figure}[h]
    \centering
        \begin{tikzpicture}
            \begin{axis}[
                grid= both,
                xlabel = {\small Rate [bpp] $\downarrow$},
                ylabel = {\small PSNR RGB [dB] $\uparrow$ } ,
                xmin = 0., xmax = 0.20,
                ymin = 24, ymax = 40,
                ylabel near ticks,
                xlabel near ticks,
                width=\linewidth,
                height=7.2cm,
                xtick distance={0.05},
                ytick distance={4.},
                xticklabel style={
                    /pgf/number format/fixed,
                    /pgf/number format/precision=2
                },
                ylabel style={yshift=-2pt},
                minor y tick num=1,
                minor x tick num=0,
                legend style={at={(1.0,0.)}, anchor=south east},
            ]
    
                \addplot[thick, gray, mark=*, mark options={solid}] table [x=rate_bpp,y=psnr_db] {data/UVG/dcvc.tsv};
                \addlegendentry{\scriptsize DCVC-FM \cite{dcvc-fm}}

                \addplot[ultra thick, myred, mark=*, mark options={solid}] table [x=rate_bpp,y=psnr_db] {data/UVG/ours.tsv};
                \addlegendentry{\scriptsize \textbf{Ours}}
    
                \addplot[thick, mygreen, mark=*, mark options={solid}] table [x=rate_bpp,y=psnr_db] {data/UVG/c3.tsv};
                \addlegendentry{\scriptsize C3 \cite{c3}}

                \addplot[thick,  myblue, mark=*, mark options={solid}] table [x=rate_bpp,y=psnr_db] {data/UVG/coolchic.tsv};
                \addlegendentry{\scriptsize Cool-chic video \cite{coolchicvideo}}

                \addplot[thick, mypurple, mark=*, mark options={solid}] table [x=rate_bpp,y=psnr_db] {data/UVG/hinerv.tsv};
                \addlegendentry{\scriptsize HiNeRV \cite{hinerv}}
    
            \end{axis}
        \end{tikzpicture}
    \caption{Random access rate-distortion graphs. UVG dataset. PSNR in RGB domain.}
    \label{fig:rd-graph-ra:uvg}
\end{figure}

Previous work \cite{coolchicvideo} performs frame-wise rate allocation using a
different rate constraint $\lambda_t$ for each frame, coupled with frame-wise RD
optimisation. With this method, each frame is encoded separately according to
its own RD cost:
\begin{equation}
    \mathcal{L}_t = 
    \mathrm{D}(\img, \sysout)
    + \lambda_t \left(\mathrm{R}(\qlatent_{t}^{m}) + \mathrm{R}(\qlatent_{t}^{r}) \right),
    \label{eq:rd-cost-frame}
\end{equation}
where $\qlatent_{t}^{m}$ and $\qlatent_{t}^{r}$ are the latent representations
respectively carrying the motion and the residue information. Yet,
determining the proper $\lambda_t$ is a difficult issue as it varies with the
the video movements, spatial complexity and bitrate.

To solve this issue, we propose an additional training step which complements
the frame-wise RD optimization. It refines all the parameters (latent
representation and neural networks) of all frames to minimize the overall video
RD-cost:
\begin{equation}
    \mathcal{L} =
    \sum_t \mathrm{D}(\mathbf{x}_t, \sysout)
    + \lambda \sum_t \mathrm{R}(\qlatent_{t}^{m}) + \mathrm{R}(\qlatent_{t}^{r}) .
    \label{eq:rd-cost-video-rewritten}
\end{equation}
This permits to have a single rate constraint $\lambda$ balancing the total
distortion of the video against the total rate. Within this overall bit budget,
the optimization process automatically allocates the rate across video frames.

\subsection{Learning more accurate motion information}
\label{sec:motion-learning}

Having a dedicated motion decoder allows to pre-train only the motion-related
parameters of the system and prevents interfering with the parameters dedicated
to the residue. We introduce a pre-training stage to first learn accurate motion
information. Then, the residue is learned without undoing the motion
representation since it has its own representation.

Estimating accurate motions is a challenging task involving elaborate strategies
\cite{DBLP:journals/pr/ZhaiXLK21}. Modern approaches
\cite{raft,uniflow,Shi_2023_CVPR} rely on 4D correlation volumes, hierarchical
refinements and ground truth motions from annotated datasets. Such pre-trained
optical flow estimators are leveraged by state-of-the-art autoencoders for video
coding \cite{dcvc-fm,DBLP:conf/cvpr/LiLL23,9941493} to obtain accurate motion
information subsequently compressed and sent to the decoder. However, none of
overfitted codecs \cite{coolchicvideo,c3, ffnerv,hinerv} exploit these optical
flow estimators.

Inspired by autoencoders, we capitalize on existing optical flow estimators by
pre-training the motion decoder to replicate motions estimated by RAFT
\cite{raft}. The optical flows
estimated by RAFT between a frame $\img$ and its references are denoted 
$\motionrefone$ and $\motionreftwo$. The encoding of
$\img$ starts by training only the motion decoder to represent the
RAFT-estimated flows. This is achieved by minimizing:
\begin{equation}
    \mathcal{L}
    =  \mathrm{D}(\motionrefone, \vtone)
    +  \mathrm{D}(\motionreftwo, \vtwo)
    + \lambda_v \left(\mathrm{R}(\vtone) + \mathrm{R}(\vtwo)\right),
\end{equation}
where $\hat{\mathbf{v}}$ denotes the compressed version of the optical flows,
$\mathrm{D}$ a distortion metric, $\mathrm{R}$ the rate and $\lambda_v$ a rate
constraint. The optical flow estimator is only used at the encoder as a
guidance, maintaining a low-complexity decoder. This frame-wise pre-training
learns accurate motions while taking their rate into account.

\begin{table}[t]
    \centering
    \caption{Decoder architectures. Layers are denoted with $k$-$i$-$o$ for
    kernel size (if any), input and output features. TConv is a transposed
    convolution with a stride of 2. Motion synthesis outputs either $m = 2$
    (P-frame) or $m = 5$ (B-frame) features.}
    \begin{tabular}{c|c|c|c|c}
        Decoder                                     & ARM                   & Upsampling            & Synthesis         & Complexity    \\
                                                    & $f_{\armparam}$       & $f_{\upparam}$        & $f_{\synthparam}$ & [MAC/pixel]   \\
        \midrule
        \midrule
        \multirow{4}{*}{\footnotesize Intra}
                                                    & Linear 24-24        & \multirow{4}{*}{\footnotesize TConv 8-1-1}      & Conv 1-7-40       & \multirow{4}{*}{\footnotesize 2292}\\
                                                    & Linear 24-24        &                                                 & Conv 1-40-3       &  \\
                                                    & Linear 24-2         &                                                 & Conv 3-3-3        &   \\
                                                    &                     &                                                 & Conv 3-3-3        &  \\
        \midrule
        \multirow{3}{*}{\footnotesize Residue}
                                                    & Linear 8-8          & \multirow{3}{*}{\footnotesize TConv 8-1-1}      & Conv 1-7-28       & \multirow{3}{*}{\footnotesize 774}\\
                                                    & Linear 8-8          &                                                 & Conv 1-28-4       & \\
                                                    & Linear 8-2          &                                                 & Conv 3-4-4        & \\
        \midrule
        \multirow{3}{*}{\footnotesize Motion}
                                                    & Linear 8-8         &   \multirow{3}{*}{\footnotesize TConv 8-1-1}     & Conv 1-7-9        & 257 (P) \\
                                                    & Linear 8-2         &                                                  & Conv 1-9-$m$      & or  \\
                                                    &                    &                                                  & Conv 3-$m$-$m$    & 473 (B)   \\
        \end{tabular}
    \label{tab:config}
\end{table}

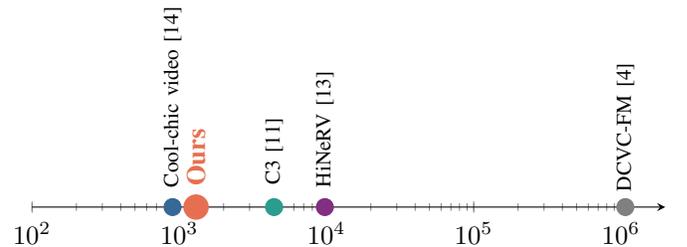
\begin{figure}[t]
    \centering
    \begin{tikzpicture}
        \begin{axis}[
            xlabel={},
            ylabel={},
            xmode=log,
            grid=none,
            ytick=\empty,
            xmin=100, xmax=2000000,
            ymin=0, ymax=10,
            axis y line=none, 
            axis x line=bottom, 
            height=4.25cm,
            width=10cm
        ]
        \addplot[only marks, mark=*, mark size=3pt] coordinates {(900, 0) (1300, 0) (4400,0) (9745,0) (1070000,0)};
        \addplot[thick, gray, mark=*, mark options={solid}, mark size=3pt] coordinates {(1070000,0)};
        \addplot[ultra thick, myred, mark=*, mark options={solid}, mark size=4pt] coordinates {(1300, 0)};
        \addplot[thick, mygreen, mark=*, mark options={solid}, mark size=3pt] coordinates {(4400,0)};
        \addplot[thick,  myblue, mark=*, mark options={solid}, mark size=3pt] coordinates {(900, 0)};
        \addplot[thick, mypurple, mark=*, mark options={solid}, mark size=3pt] coordinates {(9745,0)};
        \node[rotate=90, font=\scriptsize] at (axis cs:900,5.5) {\footnotesize Cool-chic video \cite{coolchicvideo}};
        \node[rotate=90, font=\scriptsize] at (axis cs:1300,2.5) {\normalsize \textbf{\textcolor{myred}{Ours}}};
        \node[rotate=90, font=\scriptsize] at (axis cs:4400,2.8) {\footnotesize C3 \cite{c3}};
        \node[rotate=90, font=\scriptsize] at (axis cs:9745,3.9) {\footnotesize HiNeRV \cite{hinerv}};
        \node[rotate=90, font=\scriptsize] at (axis cs:1070000,4) {\footnotesize DCVC-FM \cite{dcvc-fm}};
        \end{axis}
    \end{tikzpicture}
    \caption{Decoding complexity (MAC per decoded pixel)}
    \label{fig:complexity}
\end{figure}

\begin{figure*}[t]
    \centering
    \begin{subfigure}{0.32\linewidth}
        \centering
        \includegraphics[width=0.8\linewidth]{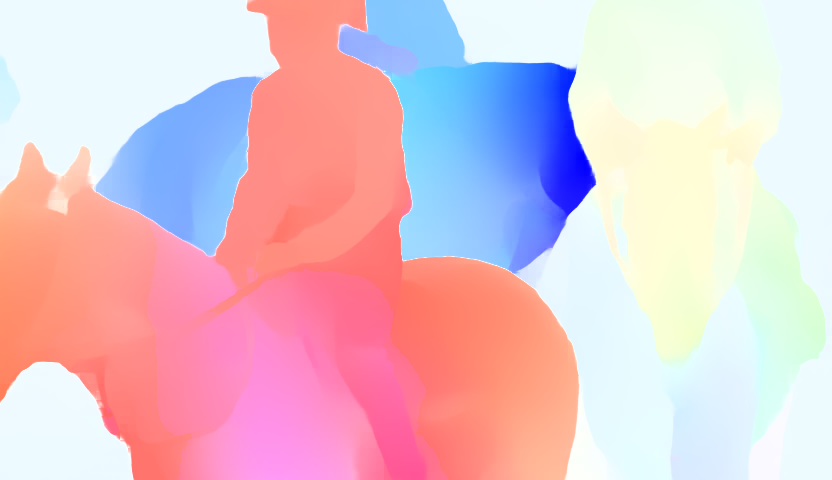}
        \caption{OF estimated by RAFT $\mathbf{v}$}
    \end{subfigure}
    \begin{subfigure}{0.32\linewidth}
        \centering
        \includegraphics[width=0.8\linewidth]{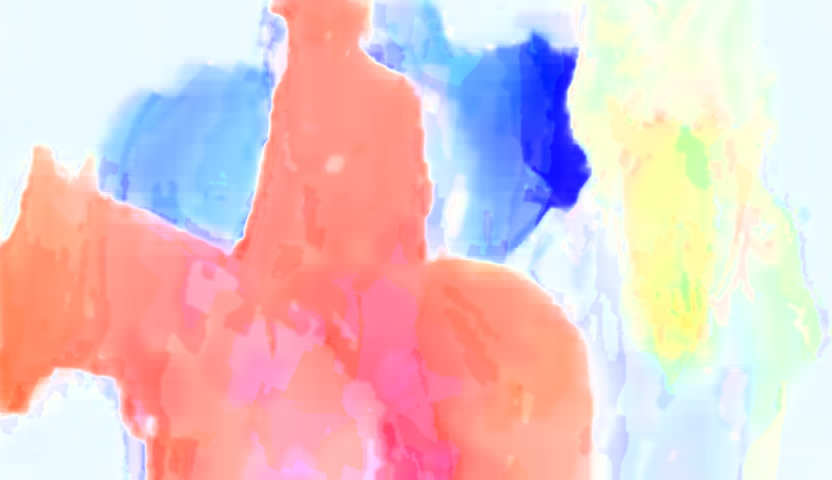}
        \caption{Decoded OF $\hat{\mathbf{v}}$ with pre-training}
    \end{subfigure}
    \begin{subfigure}{0.32\linewidth}
        \centering
        \includegraphics[width=0.8\linewidth]{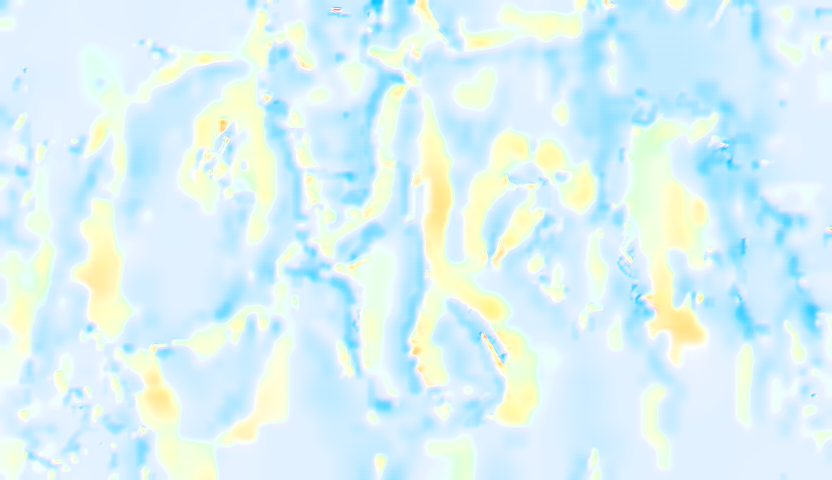}
        \caption{Decoded OF $\hat{\mathbf{v}}$ without pre-training}
    \end{subfigure}
    \caption{Importance of the optical flow (OF) pre-training for the \textit{RaceHorses} sequence (HEVC C).}
    \label{fig:flowcomp}
\end{figure*}

\section{Experiments}
\subsection{Rate-distortion performance}

The proposed method is evaluated on the HEVC and UVG datasets \cite{mpeg, uvg}
against a wide variety of other codecs: conventional codecs such as AVC
(x264-medium) and HEVC (x265-medium, HM), overfitted codecs with Cool-chic video
\cite{coolchicvideo}, C3 \cite{c3} and HiNeRV \cite{hinerv} and the autoencoder
DCVC-FM \cite{dcvc-fm}. To ensure fair comparison, conventional codecs are
evaluated in the YUV420 domain on the HEVC test sequences while overfitted
codecs and autoencoders are evaluated in the RGB domain on the UVG
sequences. Experiments are carried out on the first 9 frames of each video, with
the random access coding configuration(Fig. \ref{fig:gop:ra}).
BD-rate \cite{bdrate} (relative
rate for identical quality) is used to quantify gap performance between codecs.

Table \ref{tab:config} presents the architecture of the intra, motion and
residue decoders. Encoding starts with a frame-wise pre-training of the motion
information for $10^4$ iterations. This is followed by a joint
optimization of the
rate-distortion cost for $10^5$ iterations. Both stages schedule learning rate,
noise and softround parameters similarly to C3 \cite{c3}. The following rate
constraint $\lambda$ values are used : $\{ 0.05, 0.01, 0.0025, 0.001, 0.0005 \}$. It is also found empirically that
setting $\lambda_v = 20 \lambda$ for the motion pre-training leads to better
results.

\begin{table}[b]
    \centering
    \caption{Ablation study. BD-rates against HM on HEVC test sequences.
    Distortion is computed in the YUV420 domain.}
    \begin{tabular}{cc|cccc}
        Motion       & Joint           & \multicolumn{4}{c}{Random Access BD-rate {[}\%{]}} \\
        pre-training & optimization    & Class B     & Class C    & Class E    & Average    \\
        \midrule
        \midrule
                            & \checkmark            & 48.1        & 39.8       & 17.6       & 35.2       \\
        \checkmark          &                       & 45.7        & 61.3       & 53.8       & 53.6       \\
        \checkmark          & \checkmark            & \textbf{4.6} & \textbf{12.4} & \textbf{10.8} & \textbf{9.3}        \\
    \end{tabular}
    \label{tab:ablation}
\end{table}

Figures \ref{fig:rd-graph-ra:hevc-b} and \ref{fig:rd-graph-ra:uvg} present the rate-distortion curves obtained by the
proposed method. It significantly outperforms other overfitted codecs (C3,
Cool-chic and HiNeRV). Even in the more challenging YUV420 domain, we achieve
better compression performance than x264-medium and x265-medium and come close
to the HM. Table \ref{tab:all-bd-rate} shows the BD-rate, highlighting
the improvement offered by the proposed system against Cool-chic video. This is
particularly interesting since the main difference between these two systems is
the encoding enhancement offered by this work.

Beside its compelling compression performance, the proposed system also features
a lightweight decoder. Table \ref{tab:config} details its complexity and Fig.
\ref{fig:complexity} compares it to other overfitted and autoencoder-based
codecs. The proposed system maintains the low-complexity of Cool-chic video and
C3. It is less complex than NeRV-based codecs \textit{e.g.} HiNeRV and than
autoencoder-based codecs such as DCVC-DC.

\subsection{Ablation}

Table \ref{tab:ablation} presents the impact of removing each contribution
individually. Removing the motion decoder pre-training significantly degrades
classes B and C, but proves neutral for class E. This is due to class E
sequences having smaller movements. For sequences with substantial movements,
this pre-training enhances optical flows accuracy, as in Fig \ref{fig:flowcomp}.

\begin{figure}[t]
    \centering
    \begin{subfigure}{0.48\linewidth}
        \includegraphics[width=\linewidth]{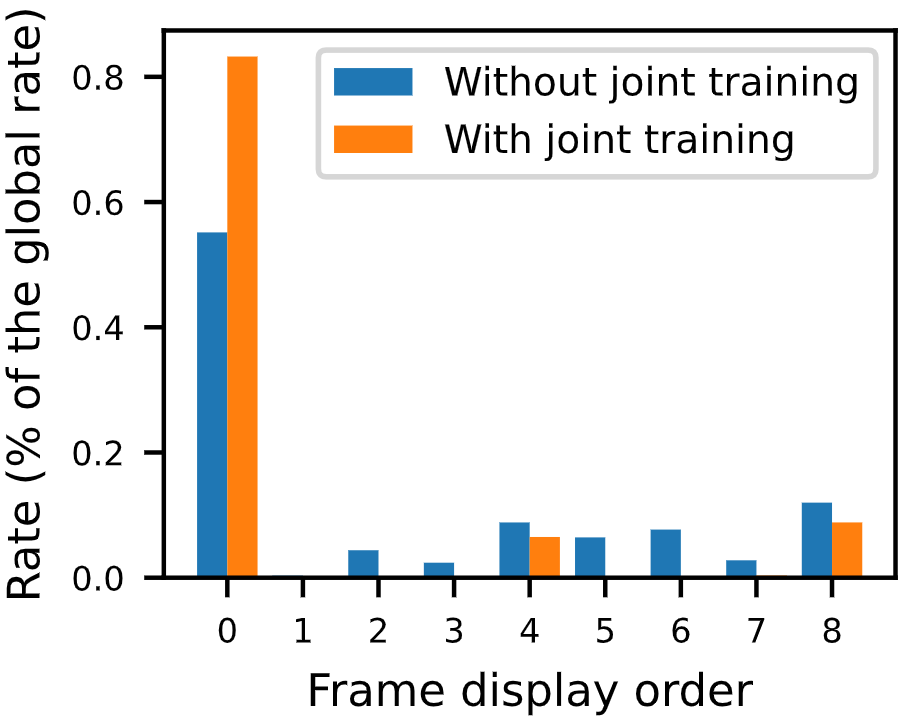}
        \caption{\textit{Four People} (HEVC E)}
    \end{subfigure}
    \begin{subfigure}{0.48\linewidth}
        \includegraphics[width=\linewidth]{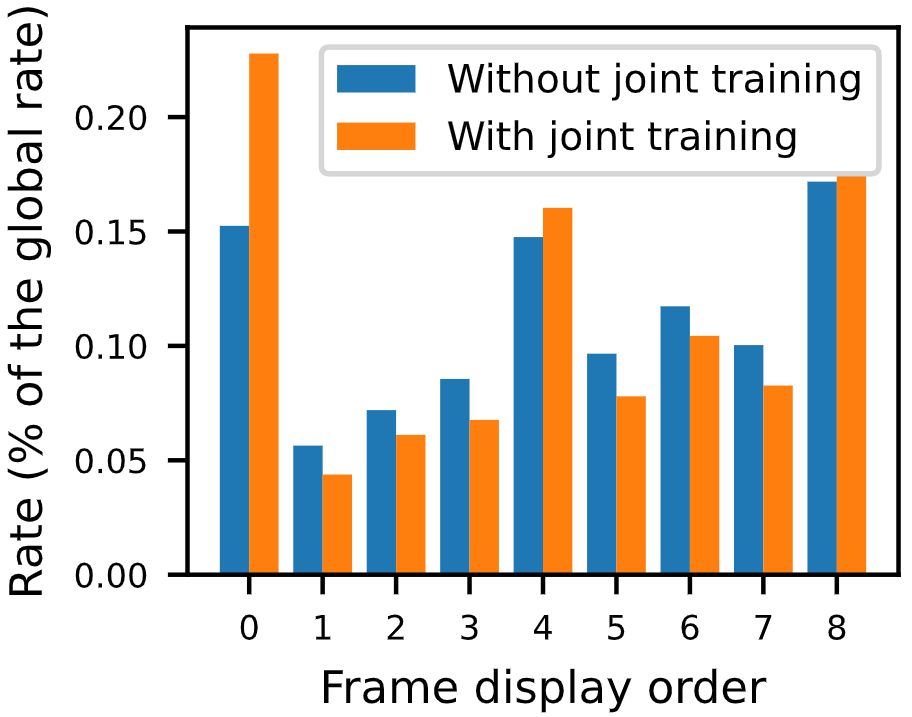}
        \caption{\textit{Ritual Dance} (HEVC B)}
    \end{subfigure}
    \caption{Rate distribution across frames.}
    \label{fig:rate_distrib}
\end{figure}

Joint optimization results in improvements for all sequences due to a more
effective rate distribution across frames. As shown in Fig
\ref{fig:rate_distrib}, the rate is more concentrated on the frequently used
frames, such as frames 0, 4, and 8. With joint optimization, the rate of the
least used frames is reduced and redistributed to frames 0, 4, and 8. This is
particularly evident in the \textit{Four People} sequence (see Figure
\ref{fig:rate_distrib}). Rather than a joint optimization, one could consider
establishing a rate constraint for each frame type, assigning a higher
constraint to less frequently used frames. However, the significant variability
in rate distributions depending on sequences complicates this approach. Joint
optimization addresses the issue of optimal rate distribution across frames
while remaining adaptable to various sequence types.

\section{Limitations}

Encoding a 1920x1080 frame lasts approximately 20 minutes on an Nvidia H100 GPU.
While this encoding time is comparable to other low-decoding complexity neural
codecs like Cool-chic video \cite{coolchicvideo} and C3 \cite{c3}, it is
significantly longer than that of autoencoder-based methods, such as DCVC-FM
\cite{dcvc-fm}, which require only a single inference for encoding. However,
recent studies indicate that the training time for overfitted codecs can be
notably reduced by trading compression performance for shorter
training\cite{blard}.

In this paper, the rate of neural network parameters is not considered during
the training process. The minimization loss (see eq.
\ref{eq:rd-cost-video-rewritten}) only focuses on the latent variable rate and
does not minimize the neural network parameters rate. Recent work proposes
methods to put the neural network rate in the training loss function, leading to
significant performance gains \cite{bull}.

\section{Conclusion}
This paper enhances overfitted codecs by improving their encoding, focusing on
motion information learning and rate allocation across frames. Key contributions
include a pre-trained motion module that boosts optical flow quality and a joint
rate-distortion optimization. Experimental results show that the proposed codec
outperforms existing overfitted codecs and competes with HEVC in the YUV420
color space while maintaining a low decoding complexity of 1300 MAC/pixel.

\newpage

\bibliography{refs}
\bibliographystyle{IEEEtran}

\end{document}